\documentclass[twocolumn]{aastex62}

\usepackage{soul}
\usepackage{amsmath}
\usepackage{subfigure}
\graphicspath{{./}{figures/}}

\received{January 1, 2018}
\revised{January 7, 2018}
\accepted{\today}

\shorttitle{Sample article}
\shortauthors{Estrela et al.}

\begin{document}

\title{The evolutionary track of H/He envelope in the observed population of sub-Neptunes and Super-Earths planets}

\correspondingauthor{Raissa Estrela}
\email{restrela@jpl.nasa.gov}

\author{Raissa Estrela}
\affil{Jet Propulsion Laboratory, California Institute of Technology \\
4800 Oak Grove Dr, Pasadena, CA 91109, USA}
\affil{Center for Radioastronomy and Astrophysics Mackenzie, Mackenzie Presbyterian University \\
R. da Consolacao, 896 - Consolacao, São Paulo - SP, 01301-000, Brazil}

\author{Mark R. Swain}
\affil{Jet Propulsion Laboratory, Caltech \\
4800 Oak Grove Dr, Pasadena, CA 91109, USA}

\author{Akash Gupta}
\affil{Department of Earth, Planetary, and Space Sciences, University of California, Los Angeles, CA 90095, USA}

\author{Christophe Sotin}
\affil{Jet Propulsion Laboratory, Caltech \\
4800 Oak Grove Dr, Pasadena, CA 91109, USA}

\author{Adriana Valio}
\affil{Center for Radioastronomy and Astrophysics Mackenzie, Mackenzie Presbyterian University \\
R. da Consolacao, 896 - Consolacao, São Paulo - SP, 01301-000, Brazil}





\begin{abstract}

The observational detection of a localized reduction in the small planet occurrence rate, sometimes termed a “gap”, is an exciting discovery because of the implications for planet evolutionary history. This gap appears to define a transition region in which sub-Neptune planets are believed to have lost their H/He envelope, potentially by photoevaporation or core powered mass loss, and have thus been transformed into bare cores terrestrial planets. Here we investigate the transition between sub-Neptunes and super-Earths using a real sample of observed small close-in planets and applying envelope evolution models of the H/He envelope together with the mass-radius diagram and a photoevaporation model. We find that photoevaporation can explain the H/He envelope loss of most super-Earths in 100Myr, although an additional loss mechanism appears necessary in some planets. We explore the possibility that these planets families have different core mass and find a continuum in the primordial population of the strongly irradiated super-Earths and the sub-Neptunes. Our analysis also shows that close-orbiting sub-Neptunes with R$<$3.5 R$_{\oplus}$ typically lose $\sim$ 30$\%$ of their primordial envelope.



\end{abstract}



\section{Introduction} \label{sec:intro}

In the last two decades we have seen a remarkable increase in the detection of exoplanets. In total, about 3946 exoplanets\footnote{https://exoplanetarchive.ipac.caltech.edu/} have been confirmed orbiting
stars with different masses, ages and spectral types (varying from M to A). Thanks largely to the NASA's {\it Kepler} mission we know today that hot-Jupiters, which were the first type of exoplanets discovered, are very rare, with an occurrence rate of only 1\% \citep{Cumming08, Howard12}. On the other hand,  small (1-4 R$_{\oplus}$), low mass ($\leq$ 20 M$_{\oplus}$) and short period (P$_{orb} <$ 100 days) planets are the most common type of exoplanet known today \cite{Benneke19}. 

Recent work by \cite{fulton2017} demonstrated the presence of a planet occurrence rate deficit, or a “gap”, between 1.5-2.0 $R_{\oplus}$, in the small and short-orbit ($\lesssim$ 100 days) population. The location of the gap can vary with stellar type, with work by \cite{Fulton18, Mc19, Wu19, gupta2019b} suggesting that the gap is centered at $\sim$ 1.6R$_{\oplus}$ for M and K dwarfs. The planet occurrence rate deficit separates the close-in planets into two categories: rocky planets with $\lesssim$1.75R$_{\oplus}$, also known as super-Earths, and planets that likely retain an H/He envelope with a size of $\gtrsim$1.75R$_{\oplus}$, redenominated sub-Neptunes \citep{owen13, lopez13}. 

A possible explanation for the "gap" is that the short-orbit sub-Neptunes could have lost their H/He envelope due to photoevaporation because of the high energy radiation from their host-stars between 20-100 Myrs. At this age, late type stars rotate faster, exhibiting higher levels of X-ray and EUV luminosity \citep{jackson12}. Numerous theoretical models \citep{owen13,Jin14,lopez13,lopez2014,Howe14,Howe15,Rogers15,owen17,Van18} show that H/He envelopes of close-orbiting sub-Neptunes can be photoevaporated. The atmospheric escape occurs in a relatively short timescale ($\sim$10$^{5}$ years), and results in a clear separation in the radius distribution between the bare cores and the planets that still hold an envelope \citep{jin2018}. This is mainly because if a planet has an atmosphere of even 1\% by mass, the presence of the envelope will have a large impact on the observed planetary radius \citep{lopez2014}.

Alternatively, atmospheric loss driven by a planet's internal luminosity, i.e., the core-powered mass-loss mechanism, can also explain the bimodal distribution of planets, even without photoevaporation \citep{Ginzburg18,gupta2019a,gupta2019b}. The source of this internal luminosity is the planet's primordial energy from formation which can be comparable or even higher than the gravitational binding energy of its atmosphere. Like photoevaporation, core-powered mass-loss can also reproduce the observed bimodal distribution in planetary occurance radii.




Analysis of the density-radius relation for small planets shows that terrestrial planets separate into two families with likely different formation histories \citep{Swain19}. One family of terrestrial planets, identified as T1, is Earth-like in composition and share a formation history compatible with the solar system. Another family of terrestrial planets, identified as T2, are denser and likely to be the remnant cores of planets that lost their primordial H/He envelopes. Here we extend the analysis from \cite{Swain19} by investigating the hypothesis whether T2 planets are bare cores of sub-Neptunes, identified as SN, under photoevaporation. Our motivation is to use a real sample of close-in planets as planets properties are influenced by multi dimensional factors (magnetic field, density, stellar wind, stellar activity (XUV luminosity), mass-radius relationship, irradiation, etc) that can be challenging to incorporate in a synthetic population. We begin exploring the plausibility of photoevaporation using a highly simplified approach to the relationship between the T2 and SN planets.  We show that this simplified approach is not adequate and use it to motivate a more physically representative approach that estimates the evolution of the H/He envelope of the observed super-Earths and sub-Neptunes using the mass-radius diagram and a photoevaporation model.

 We aim to verify if photoevaporation would be sufficient to strip the atmosphere of the T2 planets in 100Myrs and to reproduce the mass-radius diagram with the primordial and current population of SN and T2.


This paper is organised as follows. Section \ref{sec:methodsresults} provides details in how we use the mass-radius diagram to estimate the envelope fraction lost by the T2 planets and describes the photoevaporation model. Our discussions are present in Section \ref{sec:disc}, which includes an alternative method to investigate the self consistency that T2 planets are photoevaporated sub-Neptunes. Finally, Section \ref{sec:conc} concludes the paper with our main results.



\section{Methods and Results} \label{sec:style}
\label{sec:methodsresults}

\begin{figure}[h]
\begin{center}
 \includegraphics[width=0.45\textwidth]{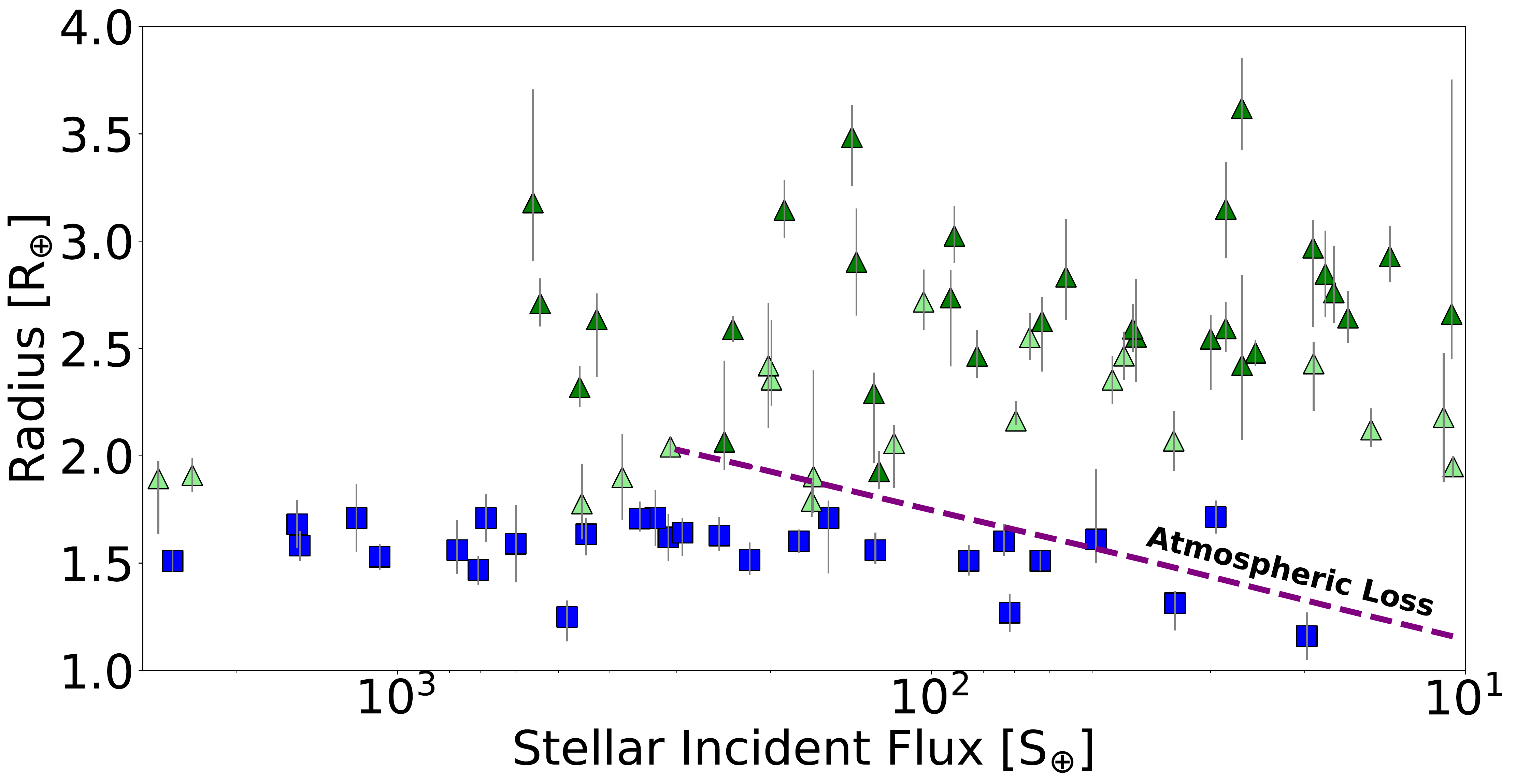} 
 \caption{High insolation terrestrial planets (T2) and sub-Neptunes (SN) in a two dimensional distribution of the planet size and the stellar insolation together with a scaling relation for the photoevaporation valley slope predicted by \cite{lopez2016}.}
   \label{fig:photslope}
\end{center}
\end{figure}

\begin{figure}[h]
\begin{center}
 \includegraphics[width=0.48\textwidth]{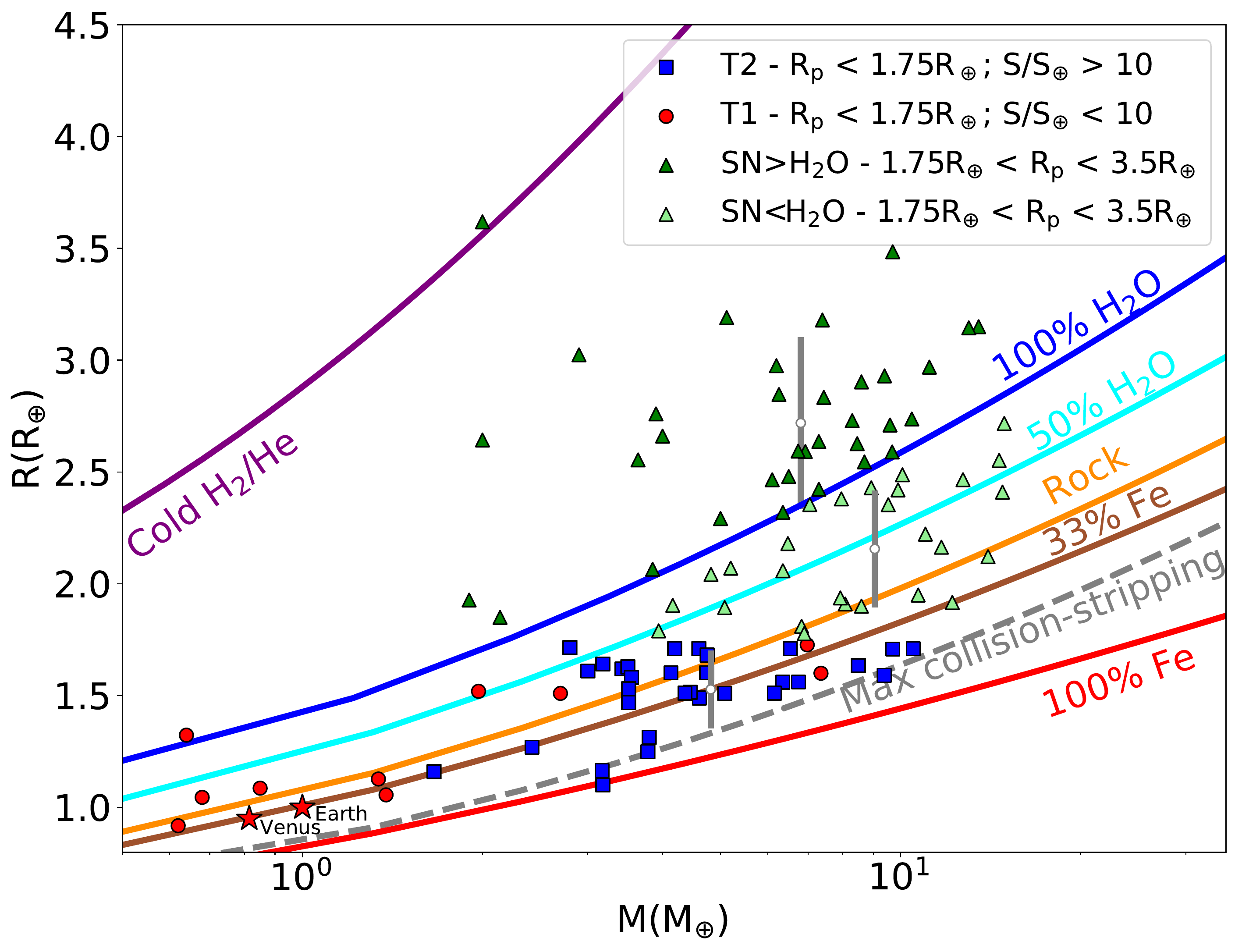} 
 \caption{Mass versus radius diagram shows the different regimes occupied by the T2 (blue), SN (dark green) and SN$<$H$_2$O (light green) planet populations. The grey symbols indicate the average mass of each population and their associated errors. SN can confidently be identified as small gas giant planets if they possess a significant H/He envelope. T2 planets can be confidently identified as envelope free (bares cores). The SN$<$H$_2$O group occupies the transition region and is thus excluded from the analysis. Low insolation T1 terrestrial planets included for context.}
   \label{fig:massrad}
\end{center}
\end{figure}

We use a sample of 100 observed close-in small planets (R$_{p}$ $<$ 3.5R$_{\oplus}$) to investigate the processes that shape sub-Neptunes into super-Earths. We base this study on the planet sample from \cite{Swain19} and also adopt their criteria based on the radius-insolation-density relationship that separates terrestrial planets into two families, T1 and T2, and separates terrestrial planets from sub-Neptunes. We display these populations in the mass-radius diagram from \cite{Zeng19}, and estimate the original envelopes of the T2 by performing the following steps:

\begin{itemize}

    \item[1.] First, as the T2 are thought to be bare core planets, we assume that T2 and SN planets both have the same underlying planet core mass distribution. All but one of the T2 planets fall below the photoevaporation valley slope \citep{lopez2016,fulton2017}, as indicated in Fig.\ref{fig:photslope}, implying that they have lost their entire envelope.
    \item[2.] The mass-radius diagram shows that the T2 planets are consistent with a rocky composition. Whereas the sub-Neptunes may be consistent with an icy composition or are likely to hold a significant H/He envelope. We divide the SN planets into two groups using the 100$\%$ H$_{\rm 2}$O composition curve in the mass-radius plane (see Fig.\ref{fig:massrad}). The SN above this curve are likely to retain an H/He envelope (SN$>$H$_2$O in Fig\ref{fig:massrad}, dark green triangles) and thus can confidently be considered sub-Neptunes. Whereas those below are in the transition region between sub-Neptunes to super-Earths. Planets in this region could be consistent with either an icy composition or have an H/He envelope (SN$<$H$_2$O, light green triangles). Therefore, for our sample of sub-Neptunes in this paper we use only the SN planets because they are more likely to possess an H/He envelope.
    \item[3.] Next, we use the average mass of the SN and of the T2 planets, which is indicated in Fig.\ref{fig:massrad} by the grey symbols. If T2 are the bare cores of sub-Neptunes, the difference in the averaged masses of these two groups is the current envelope mass of the SN (or the stripped envelope mass, in the case of the T2). We find this difference to be about $\sim$3 M$_{\oplus}$.
\end{itemize}

To check the hypothesis that the T2 and SN planets have the same core mass distribution, we estimate if the T2 could have lost an envelope of $\sim$3 M$_{\odot}$ using an envelope escape model based on photoevaporation. We increase the mass and radius of the T2 planets by the difference between the average mass and radius of these planets and the SN. If the mass loss rate is X-ray driven, we use an energy limited evaporation model from \cite{Jin14}, adopting the X-ray luminosity at 100 Myrs taken from \cite{Nunez16} according to the spectral type of the star. If the mass loss rate is in the EUV regime, we calculate the energy limited escape model using the F$_{\rm EUV}$ derived from a relation between X-ray and EUV fluxes from \cite{Chadney15} (for G and K stars) and from \cite{guinan19} (for M dwarfs). We show this calculation in greater detail in Appendix A, including the criteria to define if the planets are in the X-ray or EUV driven regime.



The results for the mass loss rate of the SN and T2 planets are shown in Figure \ref{fig:massloss} (a) in Appendix A, in which the red solid line represents the threshold for evaporating an envelope of $\sim$3M$_{\odot}$ in 100Myr. We find that the SN and T2 planets have comparable mass loss rates. In particular, the mass loss rates of the T2 planets are not sufficient to photoevaporate a 3M$_{\oplus}$ envelope in 100Myr. In accordance, theoretical model shows that a 3M$_{\oplus}$ envelope at the radii of the observed SN ($\sim$3-4 R$_{\oplus}$) corresponds to an envelope mass fraction of 5-10\% (Fig. 1 of \cite{lopez2014}) requiring total planet masses $>$30M$_{\oplus}$ , which is not in agreement with the observed SN planets.





\section{Discussion}
\label{sec:disc}

The bulk composition and the atmospheric properties of the close-in low mass exoplanet population depends on their envelope loss due to the luminosity of the host star. Assuming that T2 and SN planets have the same core masses, the difference between the averaged masses of these two groups would represent the current envelope fraction of the SN, which is found to be 3M$_{\oplus}$. This value corresponds to the minimum envelope mass that should have been lost by the T2 planets in the first 100Myrs. However, we find that photoevaporation alone is not sufficient to explain the loss of a 3M$_{\oplus}$ envelope in 100 Myrs. Therefore, this interpretation appears to raise a question whether photoevaporation can create T2 planets from the escape of the atmosphere of the sub-Neptunes or if another process is required.



\begin{figure}[!ht]
\begin{center}
 \includegraphics[width=0.45\textwidth]{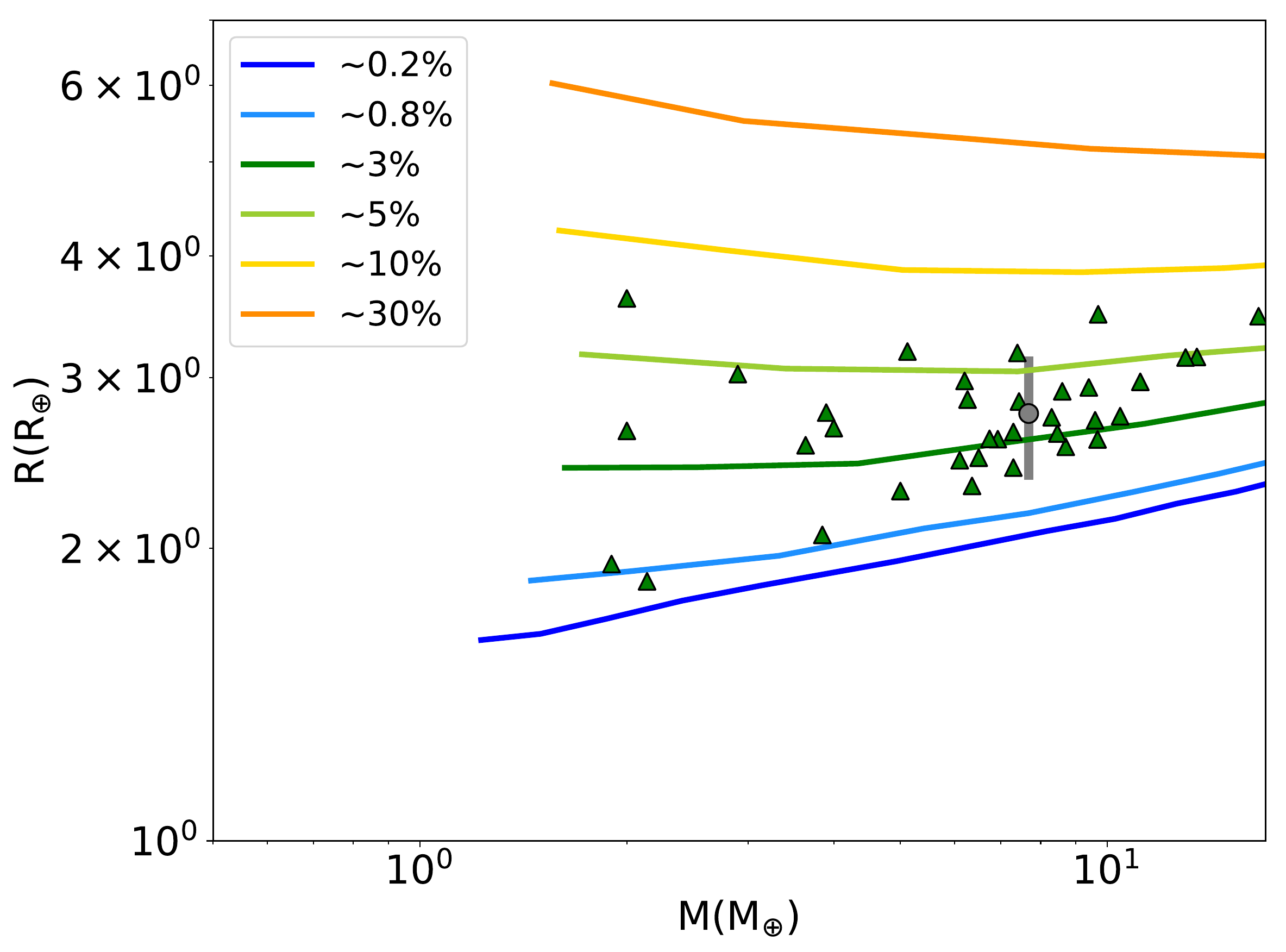}
 \caption{Mass versus radius diagram for the sub-Neptunes, SN (green) with envelope fractions from Lopez $\&$ Fortney (2014) for a planet with 100$F_{\oplus}$ and 5Gyr.}
   \label{fig:lopezfrac}
\end{center}
\end{figure}

Another possibility is that the T2 and the SN have different core masses. To explore this hypothesis, we need to determine the core mass and the primordial envelope of the SN. To estimate the core mass of SN planets, we need to know the fraction of their current envelope mass. For that, we use the envelope fractions from \cite{lopez2014} for a planet with 100F$_{\oplus}$ and 5 Gyr, which corresponds to the average insolation and age of the SN. We find that for an average mass of 6.8M$_{\oplus}$ for the SN, these planets should have an envelope fraction that corresponds to $\sim$ 4\% of their mass, as indicated in Fig. \ref{fig:lopezfrac} by a grey symbol. This implies that the current average envelope mass of the SN is 0.27 $\pm$ 0.12 M$_{\oplus}$, and the solid mass, 6.5 $\pm$ 2.9 M$_{\oplus}$, corresponds to the average mass of their core. In the case of the individuals SN, we also use the representative value of the population to find their current envelope fraction.

In the case of the T2 planets, as they are bare cores, the averaged mass of this family is given by the mass of the core, which is found by \cite{Swain19} to be 4.8 $\pm$ 1.8 M$_{\oplus}$. This value agrees well with theoretical predictions by \cite{Lee19} of 4.3 $\pm$ 1.3 M$_{\oplus}$ for the core mass of super-Earths/sub-Neptunes. 

To estimate the primordial envelope masses of each individual SN and T2 planets, we use the gas-to-core mass ratio sufficient to be accreted in $\sim$ 12Myrs given the core mass of the planet from \cite{Lee19}. We exclude 8 SN planets that have masses greater than 15M$_{\oplus}$ because \cite{Lee19} relation results in unrealistic core masses for these cases. Then, by adding this primordial envelope mass to their core mass we obtain the original masses of the planets. We find that our primordial masses are in accordance with the mass range of the initial population of \cite{Modi20} that reproduces the bimodal distribution in planetary radii after exposure to the host star's XUV irradiation. To estimate the primordial radius, we compute the radius of the primordial envelope using Eq. 4 of \cite{lopez2014} for an enhanced opacity and assuming that the planets have cores with Earth-like composition, and add it to their current radius. Table~\ref{tab:mass} shows the values obtained for the primordial mass and radius of the T2 and SN compared to their current observed values and Table~\ref{tab:mass2} shows the values estimated for the primordial envelopes.


We compute the mass loss rates of the primordial SN and T2 planets using the photoevaporation model described in the previous section. We find that the T2 and SN planets lost in average $\sim$100 $\%$ and $\sim$30$\%$, respectively, as shown in Fig.\ref{fig:envfrac}. This implies that most of the T2 planets have their envelope completely stripped. The values of the current observed masses and the mass after photoevaporation are consistent within 1$\sigma$, as shown in Table~\ref{tab:mass}.

\begin{figure*}
    \centering
    \includegraphics[width=0.6\textwidth]{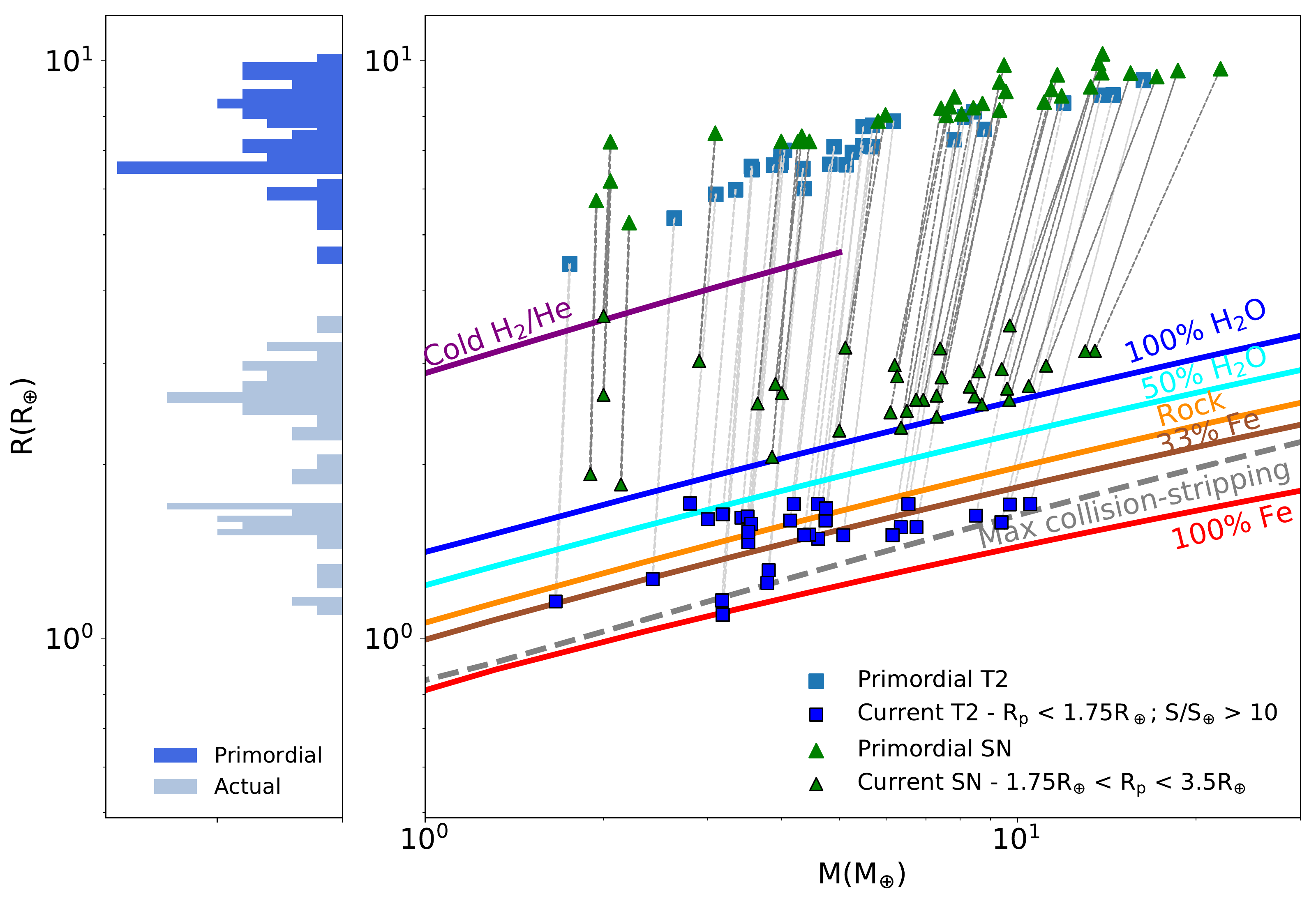} 
    \caption{Evolutionary track of the bare core super-Earths (T2) and sub-Neptunes (SN). The dashed lines connect the current planets to their primordial values obtained by adding the primordial envelope fraction. SN planets typically loose 30$\%$ of their primordial H/He envelope.}
    \label{fig:evolution}
\end{figure*}


We reach similar conclusions if the T2 planets lost their envelopes by the core-powered mass-loss mechanism \citep{Ginzburg18}. Due to their short orbital periods, the maximum initial envelope fractions of the T2 population that can be lost is given by the energy limit, which sets the maximum thermal energy that can be stored in the core and that can unbind the atmosphere \citep[see][for details]{gupta2019a}. 

\begin{figure}[!ht]
\begin{center}
 \includegraphics[width=0.45\textwidth]{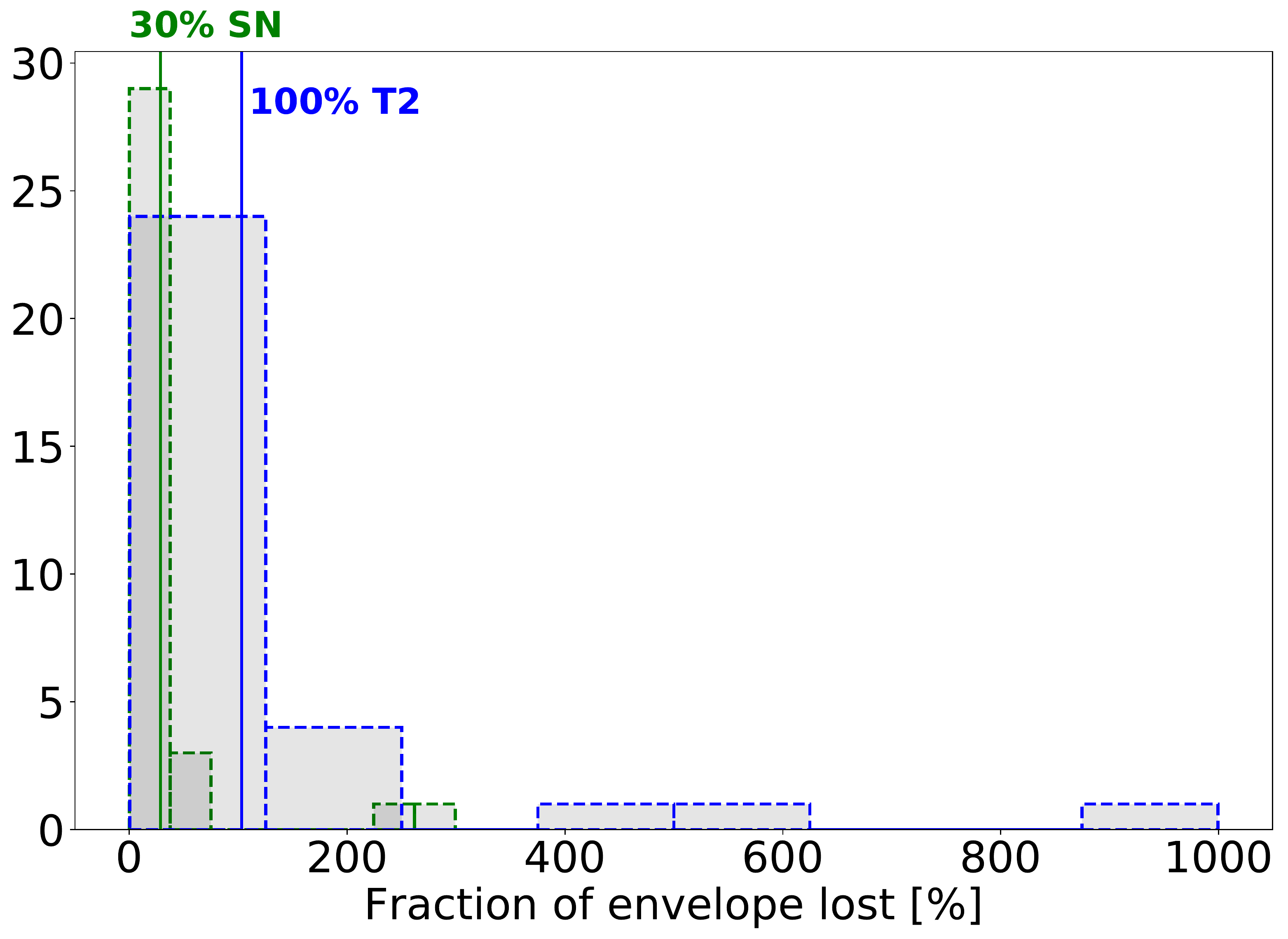}
 \caption{Fraction of the original envelope that was lost in 100 Myr by the super-Earths T2 (blue) and sub-Neptunes (green), with the median indicated by a solid line.} 
  \label{fig:envfrac}
\end{center}
\end{figure}

We build an evolutionary track of the H/He envelope of super-Earths and sub-Neptunes in the mass-radius plane. This result is shown in Figure\ref{fig:evolution}, in which the dashed lines connect the current T2 and SN to their primordial ones. The primordial T2 and SN seems to form a continuum, however SN planets have higher masses as they have higher cores, while T2 have lower masses. SN have an average radius of 8.3 $\pm$ 1.2 R$_{\oplus}$ that is compatible within 1$\sigma$ with the radius of the T2 (7 $\pm$ 1.0 R$_{\oplus}$).



Therefore, under the assumption that T2 and SN have different core masses, the photoevaporation model can explain in average the lost of the envelope of the T2 planets in 100Myr. However, if we examine the individual T2 plants, 5 out 21 of them require an additional envelope loss process to boost their mass loss rate and result in the total removal of their envelopes. This illustrate how considering envelope loss is important and how it improves the sensitivity if a bulk composition analysis is taken into consideration. While the estimates of the core mass for the T2 and SN planets formally overlap at 1$\sigma$ level, envelope loss modeling provides a compelling evidence that the cores of SN planets are indeed $\sim$35$\%$ more massive than the cores of the T2 planets.

\begin{table*}[]
\refstepcounter{table}\label{tab:mass}
\hspace{5cm}
\resizebox{0.80\textwidth}{!}{\begin{minipage}{\textwidth}
\begin{tabular}{lll}
\multicolumn{3}{c}{Table I}                                                                      \\
\multicolumn{3}{c}{Primordial/Current Observed Masses}                                                    \\
\hline
   & Primordial Mass {[}M$_{\oplus}${]}          & Observed Mass {[}M$_{\oplus}${]}              \\
T2 & 6.0 $\pm$ 3.5                               & 4.8 $\pm$ 2.2                                 \\
SN & 9.0 $\pm$ 5.0                               & 6.8 $\pm$ 3.0                                   \\
\multicolumn{3}{c}{Primordial/Current Observed Radius}                                                    \\
\hline
   & Primordial Radius  {[}R$_{\oplus}${]}       & Observed Radius {[}R$_{\oplus}${]}                              \\
T2 & 7.0 $\pm$ 1.0                               & 1.5 $\pm$ 0.2                                 \\
SN & 8.3 $\pm$ 1.2                               & 2.7 $\pm$ 0.4                                 \\
\end{tabular}
\end{minipage}}
\end{table*}

\begin{table*}[]
\refstepcounter{table}\label{tab:mass2}
\hspace{4cm}
\resizebox{0.80\textwidth}{!}{\begin{minipage}{\textwidth}
\begin{tabular}{lll}
\multicolumn{3}{c}{Table II}                                                          \\
\multicolumn{3}{c}{Primordial properties of the envelope}                             \\
\hline
   & Primordial envelope mass [M$_{\oplus}$] & Primordial envelope radius  [R$_{\oplus}$] \\
T2 & 1.3 $\pm$ 1.4                         & 5.5 $\pm$ 0.9                            \\
SN & 2.6 $\pm$ 2                           & 5.6 $\pm$ 1.0                           
\end{tabular}
\end{minipage}}
\end{table*}

\section{Conclusions}
\label{sec:conc}

When considering that the bare core super-Earths (T2) and the sub-Neptunes (SN) have the same core mass, we find that photoevaporation is not able to explain the loss of an envelope of 3M$_{\oplus}$ in 100Myrs, that represent the difference in average masses between these two populations. When we assume that super-Earths (T2) and sub-Neptunes (SN) have different core mass (SN cores are 35\% more massive), we estimate the primordial envelope for these two categories of planets and find that in average they are consistent with a typical envelope mass of 16 $\pm$ 7$\%$ and 23$\pm$ 8$\%$ of their original mass, overlapping at the 1$\sigma$ level. Also, the primordial SN and T2 have in average radius envelope fractions that overlap at 1$\sigma$ level (see Table~\ref{tab:mass}). This suggests that T2 and SN were possibly part of the same population. We visualized this in the evolutionary map (Fig. \ref{fig:evolution}) that tracks the trajectory of the current T2 and SN to their primordial values in the mass-radius plane. In terms of mass, the primordial sub-Neptunes SN have higher masses because they have higher core mass (6.5 $\pm$ 2.9 M$_{\oplus}$, in average) compared to the T2 (4.8 $\pm$ 1.8 M$_{\oplus}$). The evolutionary track also shows that in average the primordial SN planets lost $\sim$30$\%$ of their original envelopes, while the T2 lost their envelope completely. The fact that T2 are very likely to be bare cores implies that these planets are potential targets used to probe secondary atmospheres using James Webb Telescope. When considering individual planets, some of them would need an additional process (i.e core powered mass loss) to increase their mass loss rate and lead to a complete removal of their envelope in 100Myr.

\section{Acknowledgements}
We thank Hilke Schlichting for useful comments that helped to improve this manuscript. Raissa Estrela acknowledges a FAPESP fellowship ($\#$2016/25901-9 and $\#$2018/09984-7). This research was carried out at Jet Propulsion Laboratory, California Institute of Technology, under a contract with the National Aeronautics and Space Administration (80NM0018D004). © 2019 All rights reserved.



\appendix
\section{Envelope escape Model for Photoevaporation}
\label{sec:escape}

In the initial stage of planet evolution, after disk dissipation, the activity of the host star can play a significant role in shaping the atmospheric structure of a planet. In particular, young solar-type and late type stars emit large amount of EUV and X-ray radiation which decays with age as the star spin-down (\cite{Tu,jackson12}). Fully convective M dwarfs can also exhibit a rotation-activity relationship (\cite{wright18}). Either X-ray or EUV (10nm-100nm) radiation can drive the atmospheric escape depending on the location of the ionization front created by EUV flux and the sonic point in X-ray-driven flow (\cite{owen12}).

\begin{figure*}
    \centering
    \includegraphics[width=0.5\textwidth]{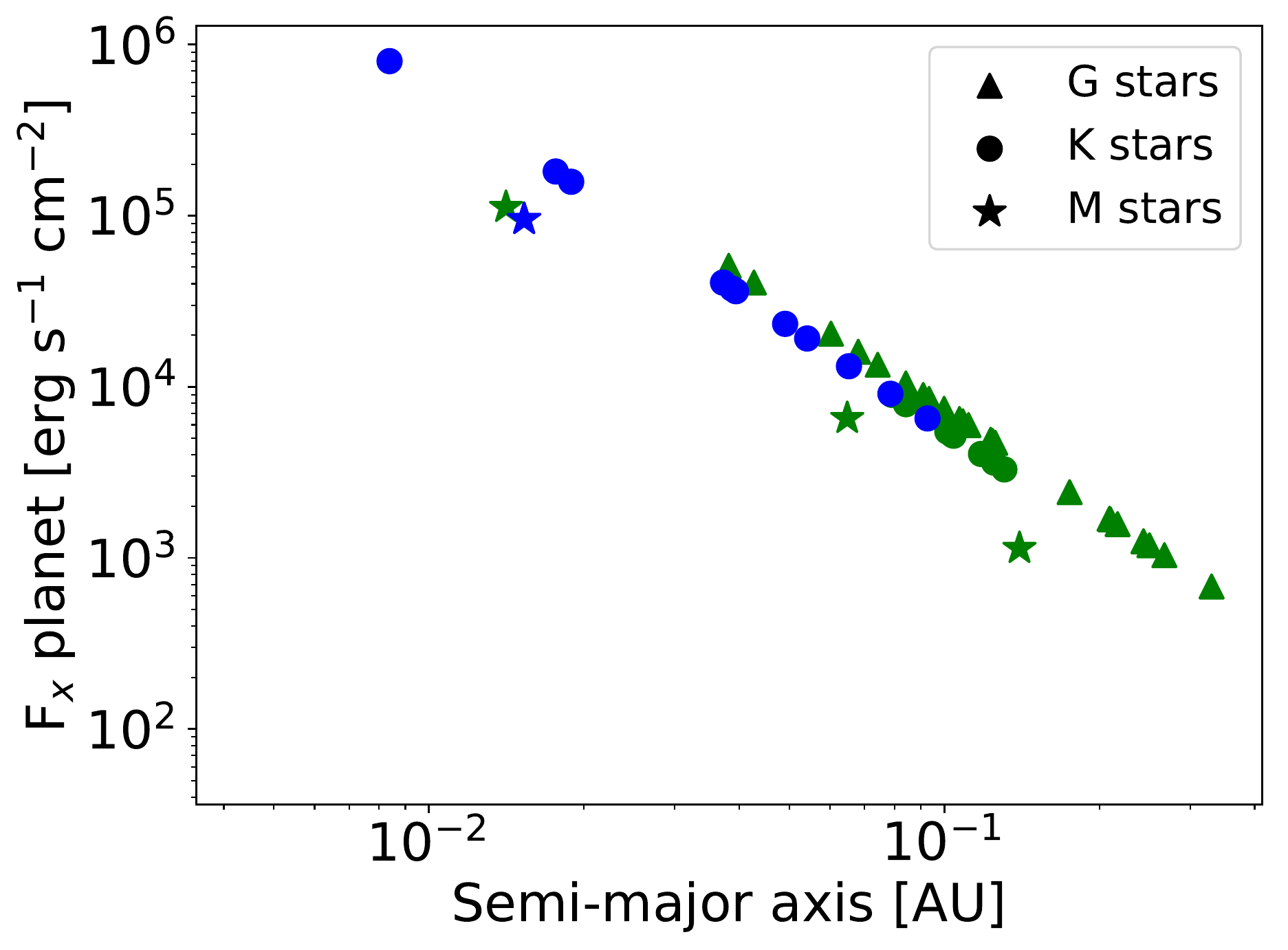} 
    \caption{F$_{EUV}$ received by the planets in our samples derived from scaling relation between X-rays and EUV fluxes at the stellar surface for each stellar spectral type.}
    \label{fig:fluxes}
\end{figure*}

X-ray driven atmospheric escape will occur mainly in the early stage of the evolution of the star. In particular, photoevaporation is likely to occur during the first 100 Myr of the planet life. Here we adopt the X-ray luminosity at 100 Myr using the correlation between X-ray luminosity and age for G and K stars from \cite{Nunez16}, while for M dwarfs we use the relation from \cite{guinan19}. Then, we calculate the X-ray fluxes received by the planets in our sample, as shown in Figure \ref{fig:fluxes}. When the X-ray luminosity falls below a critical value, atmospheric escape will transition to the EUV-driven regime.

To determine if the atmospheric escape is X-ray or EUV driven we use the criteria from \cite{owen12} which is based on the EUV luminosity of the host star. Their criteria determines the EUV luminosity necessary for the ionization front to occur before the sonic point in the X-ray flow that will make the mass loss to become EUV dominated, and is given by:

\begin{equation}
\label{eq:transition}
    \Phi_{*} \geq 10^{40} s^{-1} \Big( \frac{a}{0.1 \rm AU} \Big)^{2} \Big(\frac{\dot{m_{\rm X}}}{10^{12} g s^{-1}} \Big)^{2} \Big( \frac{A}{1/3} \Big) \times \Big( \frac{\beta}{1.5} \Big) \Big( \frac{R_{p}}{10 R_{\oplus}} \Big)^{-3}
\end{equation}

where $a$ is the semi-major axis of the planet, $\dot{m_{\rm X}}$ is the mass loss due to X-rays, $A$ is a geometry factor that takes into account how steeply the density falls off in the ionized portion of the flow, and it is typically $\approx$ $\frac{1}{3}$, and $\beta$ is an order of unity factor of the planetary radius (R$_{p}$) that characterizes the sonic surface for an X-ray dominated flow.

\subsection{EUV driven}

EUV radiation below 91.2 nm can be directly absorbed by atomic hydrogen because this is the limit that the photon has enough energy (h$\nu_{0}$ $\geq$ 13.6 eV) to ionize an H atom. Note that not all EUV can lead to hydrogen escape because some energy can be lost by radiation to space. 

At low EUV flux, the escape is mainly energy-limited. This regime assumes that a portion of the heating energy of stellar irradiation contributes to PdV work that expands the upper atmosphere (\cite{Jin14,Watson81}). The energy-limit escape occurs when the planet has an available reservoir of hydrogen. The greater is the reservoir, the more EUV energy can be absorbed to drive the escape. If the reservoir is low, the EUV energy can be either absorbed and radiated back to space or conducted to the base of the expanding thermosphere. In the energy-limited approximation, the mass loss can be estimated using the approach from \cite{Murray09}:

\begin{equation}
    \dot{m}_{\rm e-lim} = \frac{\varepsilon \pi F_{\rm EUV} R_{\rm base}^{3}}{G M_{\rm p}}
\end{equation}

where $\varepsilon$ is the heating efficiency, F$_{\rm EUV}$ is the EUV flux received by the planet, R$_{base}$ is the radius of the photoionization base, M$_p$ is the mass of the planet, and G is the gravitational constant. For both X-ray and EUV driven, we adopt a constant heating efficiency $\varepsilon$, which is a value of order 0.1 for low-mass planets \cite{owen12,owen17} . The F$_{\rm EUV}$ is calculated empirically from a power law formula of \cite{Chadney15} that describes the variation of stellar EUV flux as a function of X-ray flux.




%
%
%
%
%

\subsection{X-ray driven}

When the stars are young, they rotate faster showing a high level of activity due to an increased dynamo action, and for this reason they have a strong emission of coronal X-rays \cite{Tu}. During this stage of evolution, the orbiting planets will be mainly heated by the strong X-ray flux \cite{owen12}. Here we calculate the X-ray driven mass loss using an energy-limited model from \cite{Jin14,owen12}:

\begin{equation}
    \dot{m} = \epsilon \frac{16 \pi F_{\rm X} R_{p}^{3}}{3 G M_{p} K(\xi)}
\end{equation}

where M$_p$ is the planet mass, R$_p$ is the planet's radius at the optical depth $\tau$ = 2/3 in the thermal wavelengths, F$_{\rm X}$ is the X-ray flux calculated using the X-rays luminosity at 100Myr from \cite{Nunez16} for a G, K and M stars, $\xi$ = R$_{roche}$/R$_p$ and

\begin{equation}
K(\xi) = 1 - \frac{3}{2 \xi} + \frac{1}{2 \xi{^3}}
\end{equation}

takes into account the enhancement in the mass loss by a factor of 1/$K(\xi)$ due to the fact that the Roche lobe of a close-in planet can be close to the planet’s surface (\cite{Jin14}, \cite{Erkaev07}).

\begin{figure}
    \centering
    \subfigure[a]{\includegraphics[width=0.45\textwidth]{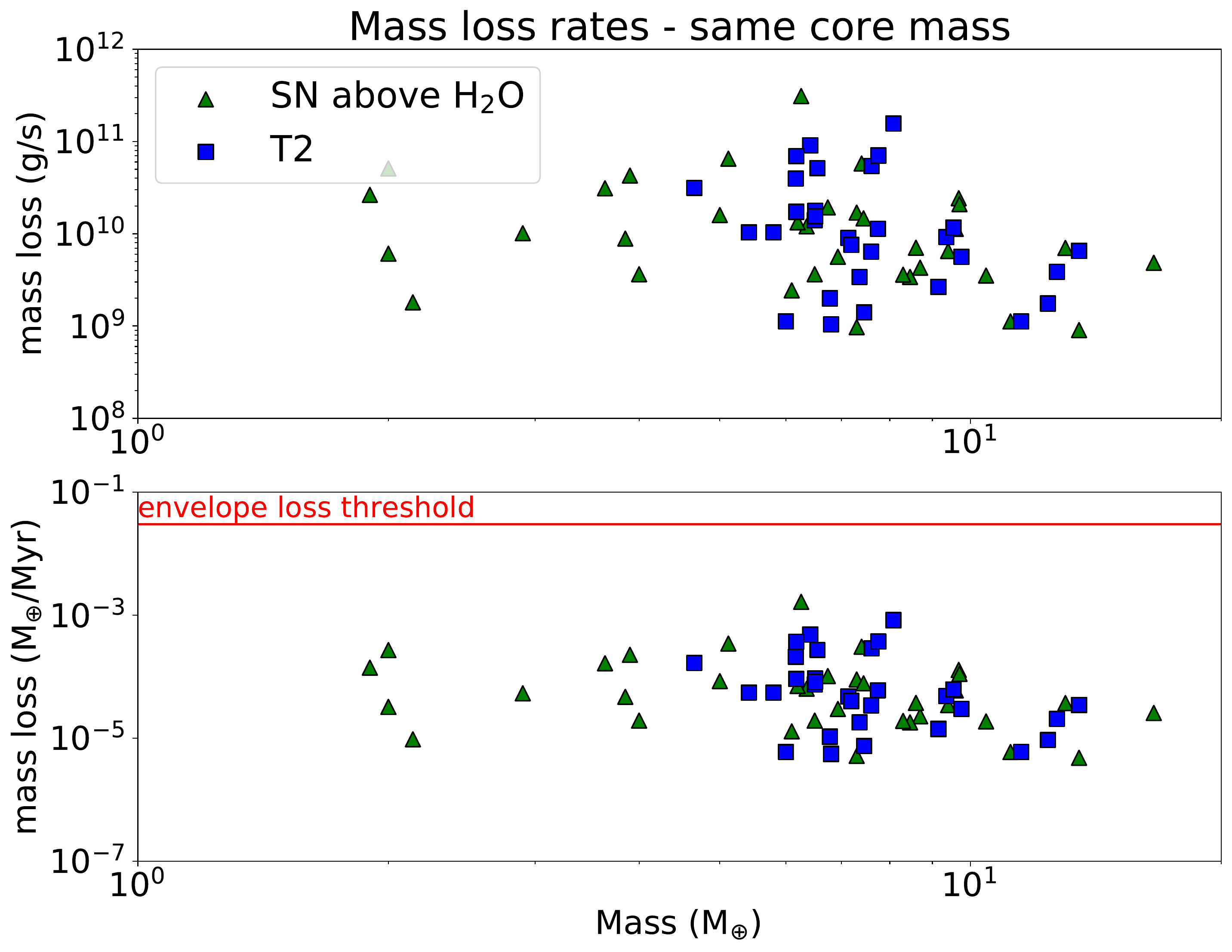}}
    \subfigure[b]{\includegraphics[width=0.45\textwidth]{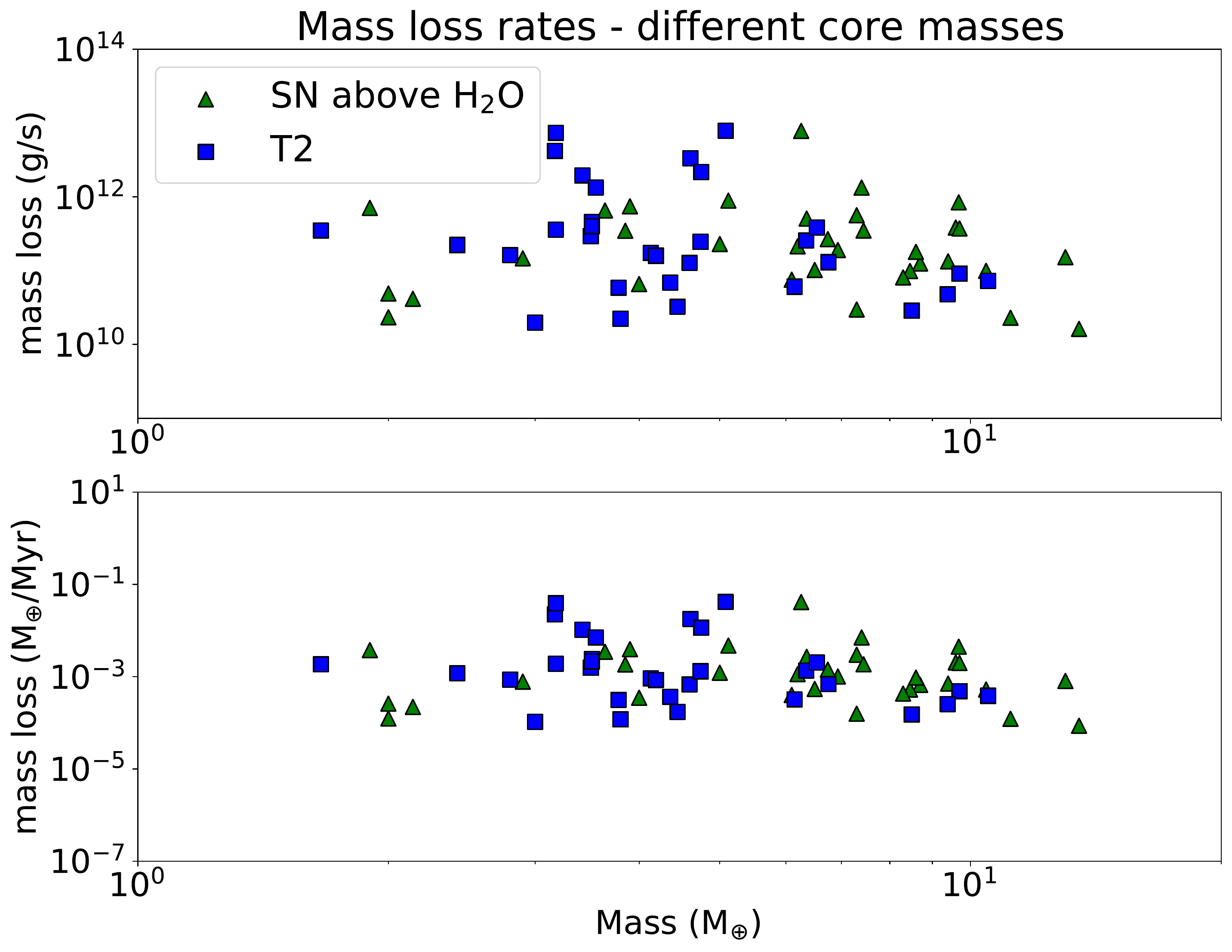}}
    \caption{[a] (\textit{Top}) Mass loss rate in g/s of the T2 (blue), SN (dark green) and SN (light green) assuming that these populations have same core masses. (\textit{Bottom}) The same but in units of M$_{\oplus}$ per Myr. The solid red line indicates the mass loss rate to lose 3M$_{\oplus}$ in 100Myr. [b] (\textit{Top}) Mass loss rate in g/s of the primordial T2 (blue), SN (dark green) and SN (light green) estimated by assuming that these population have different core masses.}
    \label{fig:massloss}
\end{figure}

Our results are shown in Fig. \ref{fig:massloss} for the mass loss rate of the SN and T2 planets in the following cases: [a] these two populations have the same core mass and the difference in the averaged masses of these two groups is the current envelope mass of the SN that is about $\sim$3 M$_{\oplus}$, the red solid line represents the threshold for evaporating $\sim$3M$_{\odot}$ in 100Myr and [b] these populations have different core masses and we use their primordial population to compute the mass loss.

\bibliographystyle{aasjournal}
\bibliography{sample62}{}

\begin{thebibliography}{}
\expandafter\ifx\csname natexlab\endcsname\relax\def\natexlab#1{#1}\fi
\providecommand{\url}[1]{\href{#1}{#1}}
\providecommand{\dodoi}[1]{doi:~\href{http://doi.org/#1}{\nolinkurl{#1}}}
\providecommand{\doeprint}[1]{\href{http://ascl.net/#1}{\nolinkurl{http://ascl.net/#1}}}
\providecommand{\doarXiv}[1]{\href{https://arxiv.org/abs/#1}{\nolinkurl{https://arxiv.org/abs/#1}}}

\bibitem[{{Benneke} {et~al.}(2019){Benneke}, {Knutson}, {Lothringer},
  {Crossfield}, {Moses}, {Morley}, {Kreidberg}, {Fulton}, {Dragomir}, {Howard},
  {Wong}, {D{\'e}sert}, {McCullough}, {Kempton}, {Fortney}, {Gilliland },
  {Deming}, \& {Kammer}}]{Benneke19}
{Benneke}, B., {Knutson}, H.~A., {Lothringer}, J., {et~al.} 2019, Nature
  Astronomy, 3, 813, \dodoi{10.1038/s41550-019-0800-5}

\bibitem[{{Chadney} {et~al.}(2015){Chadney}, {Galand}, {Unruh}, {Koskinen}, \&
  {Sanz-Forcada}}]{Chadney15}
{Chadney}, J.~M., {Galand}, M., {Unruh}, Y.~C., {Koskinen}, T.~T., \&
  {Sanz-Forcada}, J. 2015, \icarus, 250, 357,
  \dodoi{10.1016/j.icarus.2014.12.012}

\bibitem[{{Cumming} {et~al.}(2008){Cumming}, {Butler}, {Marcy}, {Vogt},
  {Wright}, \& {Fischer}}]{Cumming08}
{Cumming}, A., {Butler}, R.~P., {Marcy}, G.~W., {et~al.} 2008, \pasp, 120, 531,
  \dodoi{10.1086/588487}

\bibitem[{{Erkaev} {et~al.}(2007){Erkaev}, {Kulikov}, {Lammer}, {Selsis},
  {Langmayr}, {Jaritz}, \& {Biernat}}]{Erkaev07}
{Erkaev}, N.~V., {Kulikov}, Y.~N., {Lammer}, H., {et~al.} 2007, \aap, 472, 329,
  \dodoi{10.1051/0004-6361:20066929}

\bibitem[{{Fulton} \& {Petigura}(2018)}]{Fulton18}
{Fulton}, B.~J., \& {Petigura}, E.~A. 2018, \aj, 156, 264,
  \dodoi{10.3847/1538-3881/aae828}

\bibitem[{{Fulton} {et~al.}(2017){Fulton}, {Petigura}, {Howard}, {Isaacson},
  {Marcy}, {Cargile}, {Hebb}, {Weiss}, {Johnson}, {Morton}, {Sinukoff},
  {Crossfield}, \& {Hirsch}}]{fulton2017}
{Fulton}, B.~J., {Petigura}, E.~A., {Howard}, A.~W., {et~al.} 2017, \aj, 154,
  109, \dodoi{10.3847/1538-3881/aa80eb}

\bibitem[{{Ginzburg} {et~al.}(2018){Ginzburg}, {Schlichting}, \&
  {Sari}}]{Ginzburg18}
{Ginzburg}, S., {Schlichting}, H.~E., \& {Sari}, R. 2018, \mnras, 476, 759,
  \dodoi{10.1093/mnras/sty290}

\bibitem[{{Guinan} \& {Engle}(2019)}]{guinan19}
{Guinan}, E.~F., \& {Engle}, S.~G. 2019, Research Notes of the American
  Astronomical Society, 3, 189, \dodoi{10.3847/2515-5172/ab6086}

\bibitem[{{Gupta} \& {Schlichting}(2019{\natexlab{a}})}]{gupta2019b}
{Gupta}, A., \& {Schlichting}, H.~E. 2019{\natexlab{a}}, arXiv e-prints,
  arXiv:1907.03732.
\newblock \doarXiv{1907.03732}

\bibitem[{{Gupta} \& {Schlichting}(2019{\natexlab{b}})}]{gupta2019a}
---. 2019{\natexlab{b}}, \mnras, 487, 24, \dodoi{10.1093/mnras/stz1230}

\bibitem[{{Howard} {et~al.}(2012){Howard}, {Marcy}, {Bryson}, {Jenkins},
  {Rowe}, {Batalha}, {Borucki}, {Koch}, {Dunham}, {Gautier}, {Van Cleve},
  {Cochran}, {Latham}, {Lissauer}, {Torres}, {Brown}, {Gilliland}, {Buchhave},
  {Caldwell}, {Christensen-Dalsgaard}, {Ciardi}, {Fressin}, {Haas}, {Howell},
  {Kjeldsen}, {Seager}, {Rogers}, {Sasselov}, {Steffen}, {Basri},
  {Charbonneau}, {Christiansen}, {Clarke}, {Dupree}, {Fabrycky}, {Fischer},
  {Ford}, {Fortney}, {Tarter}, {Girouard}, {Holman}, {Johnson}, {Klaus},
  {Machalek}, {Moorhead}, {Morehead}, {Ragozzine}, {Tenenbaum}, {Twicken},
  {Quinn}, {Isaacson}, {Shporer}, {Lucas}, {Walkowicz}, {Welsh}, {Boss},
  {Devore}, {Gould}, {Smith}, {Morris}, {Prsa}, {Morton}, {Still}, {Thompson},
  {Mullally}, {Endl}, \& {MacQueen}}]{Howard12}
{Howard}, A.~W., {Marcy}, G.~W., {Bryson}, S.~T., {et~al.} 2012, \apjs, 201,
  15, \dodoi{10.1088/0067-0049/201/2/15}

\bibitem[{{Howe} \& {Burrows}(2015)}]{Howe15}
{Howe}, A.~R., \& {Burrows}, A. 2015, \apj, 808, 150,
  \dodoi{10.1088/0004-637X/808/2/150}

\bibitem[{{Howe} {et~al.}(2014){Howe}, {Burrows}, \& {Verne}}]{Howe14}
{Howe}, A.~R., {Burrows}, A., \& {Verne}, W. 2014, \apj, 787, 173,
  \dodoi{10.1088/0004-637X/787/2/173}

\bibitem[{{Jackson} {et~al.}(2012){Jackson}, {Davis}, \&
  {Wheatley}}]{jackson12}
{Jackson}, A.~P., {Davis}, T.~A., \& {Wheatley}, P.~J. 2012, \mnras, 422, 2024,
  \dodoi{10.1111/j.1365-2966.2012.20657.x}

\bibitem[{{Jin} \& {Mordasini}(2018)}]{jin2018}
{Jin}, S., \& {Mordasini}, C. 2018, \apj, 853, 163,
  \dodoi{10.3847/1538-4357/aa9f1e}

\bibitem[{{Jin} {et~al.}(2014){Jin}, {Mordasini}, {Parmentier}, {van Boekel},
  {Henning}, \& {Ji}}]{Jin14}
{Jin}, S., {Mordasini}, C., {Parmentier}, V., {et~al.} 2014, \apj, 795, 65,
  \dodoi{10.1088/0004-637X/795/1/65}

\bibitem[{{Lee}(2019)}]{Lee19}
{Lee}, E.~J. 2019, \apj, 878, 36, \dodoi{10.3847/1538-4357/ab1b40}

\bibitem[{{Lopez} \& {Fortney}(2013)}]{lopez13}
{Lopez}, E.~D., \& {Fortney}, J.~J. 2013, \apj, 776, 2,
  \dodoi{10.1088/0004-637X/776/1/2}

\bibitem[{{Lopez} \& {Fortney}(2014)}]{lopez2014}
---. 2014, \apj, 792, 1, \dodoi{10.1088/0004-637X/792/1/1}

\bibitem[{Lopez \& Rice(2016)}]{lopez2016}
Lopez, E.~D., \& Rice, K. 2016, arXiv preprint arXiv:1610.09390

\bibitem[{{McDonald} {et~al.}(2019){McDonald}, {Kreidberg}, \& {Lopez}}]{Mc19}
{McDonald}, G.~D., {Kreidberg}, L., \& {Lopez}, E. 2019, \apj, 876, 22,
  \dodoi{10.3847/1538-4357/ab1095}

\bibitem[{{Modirrousta-Galian} {et~al.}(2020){Modirrousta-Galian}, {Locci}, \&
  {Micela}}]{Modi20}
{Modirrousta-Galian}, D., {Locci}, D., \& {Micela}, G. 2020, \apj, 891, 158,
  \dodoi{10.3847/1538-4357/ab7379}

\bibitem[{{Murray-Clay} {et~al.}(2009){Murray-Clay}, {Chiang}, \&
  {Murray}}]{Murray09}
{Murray-Clay}, R.~A., {Chiang}, E.~I., \& {Murray}, N. 2009, \apj, 693, 23,
  \dodoi{10.1088/0004-637X/693/1/23}

\bibitem[{{N{\'u}{\~n}ez} \& {Ag{\"u}eros}(2016)}]{Nunez16}
{N{\'u}{\~n}ez}, A., \& {Ag{\"u}eros}, M.~A. 2016, \apj, 830, 44,
  \dodoi{10.3847/0004-637X/830/1/44}

\bibitem[{{Owen} \& {Jackson}(2012)}]{owen12}
{Owen}, J.~E., \& {Jackson}, A.~P. 2012, \mnras, 425, 2931,
  \dodoi{10.1111/j.1365-2966.2012.21481.x}

\bibitem[{{Owen} \& {Wu}(2013)}]{owen13}
{Owen}, J.~E., \& {Wu}, Y. 2013, \apj, 775, 105,
  \dodoi{10.1088/0004-637X/775/2/105}

\bibitem[{{Owen} \& {Wu}(2017)}]{owen17}
---. 2017, \apj, 847, 29, \dodoi{10.3847/1538-4357/aa890a}

\bibitem[{{Rogers}(2015)}]{Rogers15}
{Rogers}, L.~A. 2015, \apj, 801, 41, \dodoi{10.1088/0004-637X/801/1/41}

\bibitem[{{Swain} {et~al.}(2019){Swain}, {Estrela}, {Sotin}, {Roudier}, \&
  {Zellem}}]{Swain19}
{Swain}, M.~R., {Estrela}, R., {Sotin}, C., {Roudier}, G.~M., \& {Zellem},
  R.~T. 2019, \apj, 881, 117, \dodoi{10.3847/1538-4357/ab2714}

\bibitem[{{Tu} {et~al.}(2015){Tu}, {Johnstone}, {G{\"u}del}, \& {Lammer}}]{Tu}
{Tu}, L., {Johnstone}, C.~P., {G{\"u}del}, M., \& {Lammer}, H. 2015, \aap, 577,
  L3, \dodoi{10.1051/0004-6361/201526146}

\bibitem[{{Van Eylen} {et~al.}(2018){Van Eylen}, {Agentoft}, {Lundkvist},
  {Kjeldsen}, {Owen}, {Fulton}, {Petigura}, \& {Snellen}}]{Van18}
{Van Eylen}, V., {Agentoft}, C., {Lundkvist}, M.~S., {et~al.} 2018, \mnras,
  479, 4786, \dodoi{10.1093/mnras/sty1783}

\bibitem[{{Watson} {et~al.}(1981){Watson}, {Donahue}, \& {Walker}}]{Watson81}
{Watson}, A.~J., {Donahue}, T.~M., \& {Walker}, J.~C.~G. 1981, \icarus, 48,
  150, \dodoi{10.1016/0019-1035(81)90101-9}

\bibitem[{{Wright} {et~al.}(2018){Wright}, {Newton}, {Williams}, {Drake}, \&
  {Yadav}}]{wright18}
{Wright}, N.~J., {Newton}, E.~R., {Williams}, P. K.~G., {Drake}, J.~J., \&
  {Yadav}, R.~K. 2018, \mnras, 479, 2351, \dodoi{10.1093/mnras/sty1670}

\bibitem[{{Wu}(2019)}]{Wu19}
{Wu}, Y. 2019, \apj, 874, 91, \dodoi{10.3847/1538-4357/ab06f8}

\bibitem[{{Zeng} {et~al.}(2019){Zeng}, {Jacobsen}, {Sasselov}, {Petaev},
  {Vanderburg}, {Lopez-Morales}, {Perez-Mercader}, {Mattsson}, {Li}, {Heising},
  {Bonomo}, {Damasso}, {Berger}, {Cao}, {Levi}, \& {Wordsworth}}]{Zeng19}
{Zeng}, L., {Jacobsen}, S.~B., {Sasselov}, D.~D., {et~al.} 2019, Proceedings of
  the National Academy of Science, 116, 9723, \dodoi{10.1073/pnas.1812905116}

\end{thebibliography}






\end{document}